\documentclass[conference]{IEEEtran}
\IEEEoverridecommandlockouts
\usepackage[numbers, sort&compress]{natbib} % Using natbib with numerical citations and sorted, compressed numbers

\usepackage{array}
 \usepackage{booktabs}
\usepackage{mathtools}
\usepackage{microtype}
\usepackage{stfloats}
\usepackage{amsmath,amssymb,amsfonts}
\usepackage{algorithmic}
\usepackage{graphicx}
\usepackage{textcomp}
\usepackage{xcolor}
\usepackage{multicol,lipsum,multirow}
\def\BibTeX{{\rm B\kern-.05em{\sc i\kern-.025em b}\kern-.08em
    T\kern-.1667em\lower.7ex\hbox{E}\kern-.125emX}}
\begin{document}

\title{
%Beyond Speed: Unveiling the Energy Demands of ML-Driven Power Dispatch Models
Speeding Ticket: 
%The Energy and Emission Burden of ML-Accelerated Distributed Power Dispatch Models
Unveiling the Energy and Emission Burden of AI-Accelerated Distributed and Decentralized Power Dispatch Models
\thanks{This work is supported by the National Science Foundation award \#2313768 and the Air Force Office Of Scientific Research award \#FP00001854}
}

\author{\IEEEauthorblockN{Meiyi Li$^1$,  Javad Mohammadi$^1$}
\IEEEauthorblockA{\textit{$^1$Department of Civil, Architectural, and Environmental Engineering, The University of Texas at Austin}\\
meiyil@utexas.edu, javadm@utexas.edu}
}

\author{\IEEEauthorblockN{Meiyi Li$^1$, Javad Mohammadi$^1$}
\IEEEauthorblockA{\textit{$^1$Department of Civil, Architectural, and Environmental Engineering,} \\
\textit{ The University of Texas at Austin}\\
Austin, USA \\
meiyil@utexas.edu, javadm@utexas.edu}
}
\maketitle
\begin{abstract}
As the modern electrical grid shifts towards distributed systems, there is an increasing need for rapid decision-making tools. Artificial Intelligence (AI) and Machine Learning (ML) technologies are now pivotal in enhancing the efficiency of power dispatch operations, effectively overcoming the constraints of traditional optimization solvers with long computation times. However, this increased efficiency comes at a high environmental cost—escalating energy consumption and carbon emissions from computationally intensive AI/ML models. Despite their potential to transform power systems management, the environmental impact of these technologies often remains an overlooked aspect. This paper introduces the first comparison of energy demands across centralized, distributed, and decentralized ML-driven power dispatch models. We provide a detailed analysis of the energy and carbon footprint required for continuous operations on an IEEE 33 bus system, highlighting the critical trade-offs between operational efficiency and environmental sustainability. This study aims to guide future AI implementations in energy systems, ensuring they enhance not only efficiency but also prioritize ecological integrity.
\end{abstract}

\begin{IEEEkeywords}
Machine learning, Artificial intelligence, Distributed Optimization, Learning to Optimize, Power Dispatch, Carbon emission
\end{IEEEkeywords}

\section{Introduction}
The electrical grid is undergoing a transformative evolution due to the rise of decentralized generation, distributed storage, advanced technologies, and driving factors such as climate change, resilience needs, and electrification trends \cite{kargarian2016toward}. This shift results in a more distributed and interconnected system, necessitating rapid and large-scale decision-making. Traditional optimization solvers, which rely on iterative algorithms, struggle with prolonged computation times, rendering them less suitable for time-sensitive applications. To address these challenges, recent advancements in Artificial Intelligence (AI) and Machine Learning (ML) have been directed at enhancing optimization efficiency, particularly in power dispatch problems \cite{li2023hard}. Several studies have demonstrated that neural networks can drastically reduce the time required to find optimal solutions by speeding up the online search process \cite{amos2022meta}. Furthermore, the repetitive nature of power dispatch problems provides abundant historical data, which, when harnessed through ML, facilitates significant offline computation to augment the efficiency of real-time operations.

However, while the achievements in speed and efficiency are celebrated, a critical issue often remains overshadowed: the escalating energy consumption and carbon emissions associated with these computationally intensive AI/ML models. Remarkably, the computational power required to sustain the rise of AI is doubling approximately every 100 days \cite{zhu2023intelligent}. It is projected that AI workloads will account for 15\% to 20\% of total data center energy consumption by 2028 \cite{avelar2023ai}. This pursuit of high-speed computation comes with a substantial environmental cost, both in terms of energy requirements and carbon emissions, potentially negating the environmental benefits these technologies aim to provide.

Efforts have been initiated within the open-source community to develop tools that automatically measure the environmental footprint of AI/ML models, such as Code Carbon \cite{schmidt2021codecarbon} and LLM Carbon \cite{faiz2023llmcarbon}. Several key factors, including the geographic location of the server, the type of GPU, and the duration of the training, are considered to estimate the approximate CO2 equivalent emissions produced \cite{lacoste2019quantifying}. The extent of CO2 emissions associated with training models for various tasks such as Image Classification, Machine Translation, Named Entity Recognition, Question Answering, and Object Detection has been analyzed \cite{luccioni2023counting}. A systematic comparison of the on-line inference costs of various categories of ML systems is presented in \cite{luccioni2023power}, covering both task-specific (i.e., fine-tuned models performing a single task) and general-purpose models (i.e., those trained for multiple tasks). These initiatives have led to a repository for reporting and tracking energy usage and carbon emissions of models for various tasks. Nevertheless, estimating the energy and emissions impact of AI/ML models remains a relatively under-explored topic.
\begin{figure*}[htbp]
\centering
\setlength{\abovecaptionskip}{0.2cm}
\includegraphics[width=2\columnwidth]{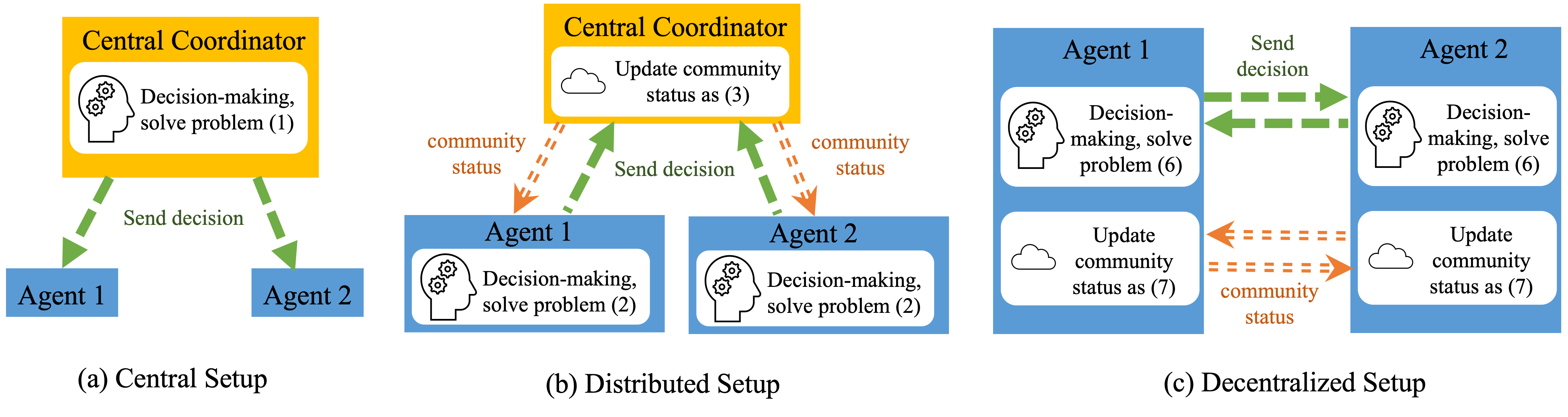}
\caption{Different power dispatch setups. In centralized decision-making, a central coordinator handles all computations and decisions, directing unidirectional communication to all nodes. Distributed decision-making involves local controllers coordinating with a central entity through bidirectional communication, while decentralized decision-making operates without a central coordinator, relying on peer-to-peer communication for independent, cooperative decisions.} 
\label{f:setup}
\end{figure*}

With the introduction of the Smart Grid, characterized by a more distributed physical structure, there has been a growing interest in distributed (decentralized) decision-making tools \cite{molzahn2017survey}. These tools are noted for their superior performance in terms of scalability, security, privacy, resilience to system disruptions, and adaptability \cite{li2023machine}. In a distributed (decentralized) setup, multiple agents—comprising controllers, devices, or algorithms—manage their specific local tasks independently while interacting or collaborating with others to achieve collective goals (see Figure \ref{f:setup}). Many machine learning-enabled solvers have been proposed, mimicking popular distributed optimization methods such as the Alternating Direction Method of Multipliers (ADMM) \cite{li2023machine,li2023learning,biagioni2020learning,mak2023learning}. In these models, the decision-making process is divided, and each agent operates autonomously, contributing to the overall system's performance.

In this paper, we present the first-of-its-kind comparison of the energy demands of ML-driven power dispatch models in centralized, distributed, and decentralized setups. We measure deployment costs in terms of the energy and carbon required to perform a full week's worth (24x7, every 15 minutes) of on-line inferences on the IEEE 33 bus system using these models. By providing a detailed account of the widely varying energy requirements for ML-driven power dispatch models, we aim to offer insights that can help practitioners better understand accuracy-efficiency trade-offs across different setups. Furthermore, our study aims to improve the precision of estimates and policy decisions within the energy sector.

\section{Problem Formulation}

This section introduces the notations and formulations of the power system energy supply-demand problem, commonly referred to as power dispatch optimization. This problem aims to identify the most cost-effective power production strategies to meet end-user demands. We present detailed formulations for centralized, distributed, and decentralized setups in the subsequent subsections.

\subsection{Centralized Formulation}

In the centralized formulation, we consider a power dispatch problem within a community of \(N_{\texttt{A}}\) agents. Each agent, indexed by \(i\) where \(i=1,...,N_{\texttt{A}}\), controls \(N_{\texttt{G}}\) generators (indexed by \(g\)) and \(N_{\texttt{D}}\) loads (indexed by \(d\)). The objective and constraints of the centralized power dispatch problem are defined as \eqref{cen problem}.

In \eqref{cen problem}, $f^{i,g}$ is the cost function of generator $g$ owned by agent $i$. $P_{\texttt{G}}^{i,g}$ and $P_{\texttt{D}}^{i,d}$ represent the power generated and consumed. Equation \eqref{cen inequality} limits the generation limits. Equation \eqref{cen po} defines the output power of each agent and equation \eqref{cen equality} maintains the supply-demand balance.

A central coordinator oversees data gathering, computation, and decision-making, ensuring efficient management and unidirectional communication from the center to all agents, as shown in Figure \ref{f:setup}.

\begin{subequations}
\label{cen problem}
\begin{gather}
    \min f= \sum_{i=1}^{N_{\texttt{A}}} \sum_{g=1}^{N_{\texttt{G}}}f^{i,g}(P_{\texttt{G}}^{i,g})\label{cen obj}\\
    \textup{s.t. }0\leq P_{\texttt{G}}^{i,g}\leq P_{\texttt{Gmax}}^{i,g},\forall g=1,...,N_{\texttt{G}}, i=1,..,N_{\texttt{A}} \label{cen inequality}\\
P_{\texttt{o}}^{i}=\sum_{g=1}^{N_{\texttt{G}}}P_{\texttt{G}}^{i,g}- \sum_{d=1}^{N_{\texttt{D}}}P_{\texttt{D}}^{i,d}\label{cen po}\\
     \sum_{i=1}^{N_{\texttt{A}}} P_{\texttt{o}}^{i}=0\label{cen equality}  
\end{gather}
\end{subequations}

\subsection{Distributed Formulation}

The distributed formulation adapts the centralized problem (\eqref{cen problem}) using the ADMM algorithm, where multiple agents operate independently but coordinate decisions through a central coordinator, allowing bidirectional communication. Each agent \(i\) solves the local problem \eqref{dis agent} in each iteration \(k\) of the communication, where $\rho>0$ is a penalty parameter.

\begin{gather}
     [P_{\texttt{G}}^{i,g,[k+1]}, g=1,..., N_{\texttt{G}}], P_{\texttt{o}}^{i,[k+1]}= \arg \min \sum_{g=1}^{N_{\texttt{G}}}f^{i,g}(P_{\texttt{G}}^{i,g})\nonumber\\
     + \lambda^{[k]}(P_{\texttt{o}}^{i} +\sum_{j\neq i} P_{\texttt{o}}^{j,[k]})
     +\frac{\rho}{2}\left (P_{\texttt{o}}^{i} +\sum_{j\neq i} P_{\texttt{o}}^{j,[k]}\right )^2 \label{dis agent}
    \end{gather}

The central coordinator updates the Lagrangian multipliers to estimate the community status as follows:

\begin{gather}
\lambda^{[k+1]}=\lambda^{[k]}+\rho(\sum_{i=1}^{N_{\texttt{A}}} P_{\texttt{o}}^{i,[k+1]} ) 
\end{gather}

\subsection{Decentralized Formulation}

In the decentralized setup, there is no central coordinator. Instead, each agent independently makes decisions, relying on peer-to-peer communications to optimize the system cooperatively. Each agent keeps a copy of the global constraints to track their neighbors' activities, denoted as \(P_{\texttt{oCopy}}^{i,j}\), for all \(j \neq i\). The original centralized constraint (\eqref{cen equality}) is replaced by the following decentralized constraints:

\begin{gather}  
P_{\texttt{o}}^{i}+\sum_{j\neq i} P_{\texttt{oCopy}}^{i,j} =0, \forall i=1,...,N_{\texttt{A}} \label{copy}\\
P_{\texttt{oCopy}}^{i,j}=P_{\texttt{o}}^{j}, \forall j\neq i,i=1,...,N_{\texttt{A}}\label{copy equality}
\end{gather}

During each communication iteration \(k\), every agent \(i\) solves a primal subproblem and updates the dual variables as described in  \eqref{decen primal} and \eqref{decen dual}.

\begin{figure*}[b]
\begin{gather}
 \textup{primal optimization: }  [P_{\texttt{G}}^{i,g,[k+1]}, g=1,..., N_{\texttt{G}}],  [P_{\texttt{o}}^{i,[k+1]}, P_{\texttt{oCopy}}^{i,j,[k+1]},\forall j\neq i] = \arg \min \sum_{g=1}^{N_{\texttt{G}}}f^{i,g}(P_{\texttt{G}}^{i,g})
    \nonumber\\ 
    +\left ( \lambda^{i,j,[k]}(P_{\texttt{oCopy}}^{i,j}-P_{\texttt{o}}^{j,[k]}) + \lambda^{j,i,[k]}(P_{\texttt{oCopy}}^{j,i,[k]}-P_{\texttt{o}}^{i})+\frac{\rho}{2} \left (  P_{\texttt{oCopy}}^{i,j}-P_{\texttt{o}}^{j,[k]}\right )^2+\frac{\rho}{2}\left ( P_{\texttt{oCopy}}^{j,i,[k]}-P_{\texttt{o}}^{i} \right )^2\right ) \label{decen primal}\\
     \textup{dual update:}\lambda^{i,j,[k+1]}=\lambda^{i,j,[k]}+\rho(P_{\texttt{oCopy}}^{i,j,[k+1]}-P_{\texttt{o}}^{j,[k+1]}),\forall j\neq i\label{decen dual}    
\end{gather}
\end{figure*}

\section{Methodology}
This section describes the use of ML-driven power dispatch models to replace traditional iterative solvers typically employed in power dispatch problems. Unlike traditional methods that require several iterations to solve optimization problems (\eqref{cen problem}, \eqref{dis agent}, and \eqref{decen primal}), our approach uses neural network approximators to map inputs directly to optimal solutions in a single feed-forward pass. This significantly accelerates computation times. We detail the application of these models in centralized, distributed, and decentralized setups, summarized in Table \ref{tab:summary}.

\subsection{Centralized Setup}
In a centralized setup, a single neural network approximator, managed by a central coordinator, processes decisions for the entire system. The network receives power consumption data from all agents and loads, denoted as \(P_{\texttt{D}}^{i,d}\), for each agent \(i = 1, \ldots, N_{\texttt{A}}\) and load \(d = 1, \ldots, N_{\texttt{D}}\). The output includes the power generation profile \(P_{\texttt{o}}^{i}\) and \(P_{\texttt{G}}^{i,g}\) for all agents \(i\) and generators \(g\). The input dimensionality is \(N_{\texttt{D}}N_{\texttt{A}}\), and the output dimensionality is \((N_{\texttt{G}}+1)N_{\texttt{A}}\).

\subsection{Distributed Setup}
In the distributed setup, each agent \(i\) employs its own neural network to make local decisions in coordination with a central coordinator. The input to the local neural network includes:

1) Local load power consumption, \(P_{\texttt{D}}^{i,d}\), for all \(d\).

2) The latest Lagrangian multiplier, \(\lambda^{[k]}\), received from the central coordinator.

3) Coupled variables from neighboring agents, \(P_{\texttt{o}}^{j,[k]}\), also received from the central coordinator.

The output is the local power generation profile, represented as $P_{\texttt{o}}^{i}$ and $P_{\texttt{G}}^{i,g}$ for all generators $g = 1, \ldots, N_{\texttt{G}}$.  The input dimension is \(N_{\texttt{D}} + N_{\texttt{A}}\), and the output dimension is \(N_{\texttt{G}} + 1\).

\subsection{Decentralized Setup}
The decentralized setup operates without a central coordinator, with each agent \(i\) independently using its own neural network based on peer-to-peer communications. Inputs to the network include:

1) Local load power consumption, \(P_{\texttt{D}}^{i,d}\),

2) Lagrangian multipliers from neighbors, \(\lambda^{j,i,[k]}\),

3) Neighbors' coupled variables, \(P_{\texttt{o}}^{j,[k]}\), and

4) Output power tracking copies from neighbors, \(P_{\texttt{oCopy}}^{j,i,[k]}\).

The network outputs:

1) The agent's power generation profile, \(P_{\texttt{o}}^{i,[k+1]}\) and \(P_{\texttt{G}}^{i,g,[k+1]}\), and

2) Tracking copies of neighbors' output power, \(P_{\texttt{oCopy}}^{i,j,[k+1]}\).

The input dimension is \(N_{\texttt{D}} + 3(N_{\texttt{A}}-1)\), reflecting the complexity of interactions, and the output dimension is \(N_{\texttt{G}} + N_{\texttt{A}}\).

\section{Experimental Results}

To assess the energy consumption and carbon emissions of deploying various models, we utilized Code Carbon \cite{schmidt2021codecarbon}, which leverages real-time monitoring and emission calculations based on local grid data. This tool measures power usage every 15 seconds using hardware-level libraries and calculates carbon emissions based on real-time electricity grid data, specifically using the average emissions intensity for Texas, USA. Our measurements spanned a comprehensive period, capturing one week's worth of inferences, computed as inferences occurring four times per hour, 24 hours a day, for seven days.

\begin{table*}[htbp]
\caption{Summary of ML-Driven Power Dispatch Models: The table details the number of neural networks utilized, input and output variables, and their respective dimensions for each setup.}
    \label{tab:summary}
    \centering
    \begin{tabular}{|>{\centering\arraybackslash}p{1.5cm}|>{\centering\arraybackslash}p{1.2cm}|>{\centering\arraybackslash}p{4.5cm}|>{\centering\arraybackslash}p{1.4cm}|>{\centering\arraybackslash}p{5cm}|>{\centering\arraybackslash}p{1.5cm}|}
        \hline
        \textbf{Setup} & \textbf{Network Number} & \textbf{Input Variables} & \textbf{Input Dim} & \textbf{Output Variables} & \textbf{Output Dim} \\
        \hline
        \textbf{Centralized} & 1 & $P_{\texttt{D}}^{i,d}, \forall i=1,..,N_{\texttt{A}}$, $\forall d=1,..,N_{\texttt{D}}$ & $N_{\texttt{D}}N_{\texttt{A}}$ & $P_{\texttt{o}}^{i}, P_{\texttt{G}}^{i,g}, \forall i=1,..,N_{\texttt{A}}, \forall g=1,..,N_{\texttt{G}}$ & $(N_{\texttt{G}}+1)N_{\texttt{A}}$ \\
        \hline
        \textbf{Distributed} & $N_{\texttt{A}}$ & 
        \begin{itemize}
            \item $P_{\texttt{D}}^{i,d}, \forall d=1,..,N_{\texttt{D}}$
            \item $\lambda^{[k]}$
            \item $P_{\texttt{o}}^{j,[k]}, \forall j \neq i$
        \end{itemize} & $N_{\texttt{D}} + N_{\texttt{A}}$ & $P_{\texttt{o}}^{i}, P_{\texttt{G}}^{i,g}, \forall g=1,..,N_{\texttt{G}}$ & $(N_{\texttt{G}}+1)$ \\
        \hline
        \textbf{Decentralized} & $N_{\texttt{A}}$ & 
        \begin{itemize}
            \item $P_{\texttt{D}}^{i,d}, \forall d=1,..,N_{\texttt{D}}$
            \item $\lambda^{j,i,[k]}, \forall j \neq i$
            \item $P_{\texttt{o}}^{j,[k]}, \forall j \neq i$
            \item $P_{\texttt{oCopy}}^{j,i,[k]}$
        \end{itemize} & $N_{\texttt{D}} + 3(N_{\texttt{A}}-1)$ & 
        \begin{itemize}
            \item $P_{\texttt{o}}^{i,[k+1]}, P_{\texttt{G}}^{i,g,[k+1]}, \forall g=1,..,N_{\texttt{G}}$
            \item $P_{\texttt{oCopy}}^{i,j,[k+1]}, \forall j \neq i$
        \end{itemize} & $N_{\texttt{G}} + N_{\texttt{A}}$ \\
        \hline
    \end{tabular}
\end{table*}

\subsection{Experimental Setup}

System: We used the IEEE 33-bus system dataset, available through MATPOWER \cite{zimmerman2010matpower}. In our configuration, each of the 33 agents managed one bus, including its associated generators and loads. To mimic real-world variability, we introduced a 10\% fluctuation at each load node.

Neural Network Architecture: All models employed fully connected neural networks with a standardized architecture to ensure consistency in our comparative analysis. The architecture was fixed with one hidden layer, where the number of hidden nodes was adjustable. The input and output dimensions were determined by the number of load and generator nodes in the community ($N_{\texttt{D}}$ and $N_{\texttt{G}}$ respectively), allowing us to control the model size through the number of hidden nodes.

Metrics: We conducted 20 simulations and reported the average results for carbon emissions (in grams of $CO_2$ equivalent) and energy usage (in kWh) across the community of all 33 agents. In distributed and decentralized setups, total energy usage and emissions were computed collectively for all agents per communication iteration.

\subsection{Results on Different Model Setups}

\textbf{Discussion:} Even for a single communication iteration, the energy consumption in distributed and decentralized setups is higher than in the centralized model.

We investigated the carbon emissions and energy usage across various model setups. Each neural network architecture was implemented with a single hidden layer, maintaining a comparable number of total parameters across the different setups. The community configuration consisted of 3300 generators and 3300 loads. After conducting 20 simulations, we analyzed the mean results, as presented in Table \ref{tab:comparison}.

\begin{table}[h!]
    \centering
    \caption{Comparative analysis of total energy usage and carbon emissions across different model setups, highlighting that even for one communication iteration, distributed and decentralized setups consume more energy than the centralized model.}
    \label{tab:comparison}
        \begin{tabular}{|c|c|c|c|}
\hline
\textbf{Setup}                                                               & \textbf{Centralized} & \textbf{Distributed} & \textbf{Decentralized} \\ \hline
\textbf{\begin{tabular}[c]{@{}c@{}}Hidden\\  nodes\end{tabular}}             & 6,000                & 5,133                & 3,655                  \\ \hline
\textbf{\begin{tabular}[c]{@{}c@{}}Total \\ parameters\end{tabular}}         & 39,807,333           & 39,809,748           & 39,807,339             \\ \hline
\textbf{\begin{tabular}[c]{@{}c@{}}Energy usage \\ (kWh)\end{tabular}}       & 0.000313             & 0.000919             & 0.001049               \\ \hline
\textbf{\begin{tabular}[c]{@{}c@{}}Carbon emission\\ (g CO2eq)\end{tabular}} & 0.000149             & 0.000438             & 0.000499               \\ \hline
\end{tabular}
\end{table}

Our findings indicate that energy consumption and carbon emissions are significantly higher in both distributed and decentralized setups compared to the centralized configuration, even though all setups have the same total number of parameters. Notably, even after a single communication iteration, the distributed and decentralized models consume more energy and emit more carbon than the centralized model. Typically, distributed and decentralized models require multiple communication rounds to achieve consensus, which further increases their energy demands.

The increased energy usage in distributed and decentralized models may be attributed to:
1) Larger neural networks in the centralized model can utilize hardware resources more efficiently, effectively leveraging the parallel processing capabilities of modern hardware.
2) Each inference operation in distributed and decentralized setups involves some fixed overhead, such as loading model parameters and initializing certain operations. This overhead is incurred once for each agent, resulting in multiplied overhead that leads to higher energy usage and carbon emissions.

\subsection{Results on Different Model Sizes}

\begin{figure*}[htbp]
\centering
\setlength{\abovecaptionskip}{0.2cm}
\includegraphics[width=2\columnwidth]{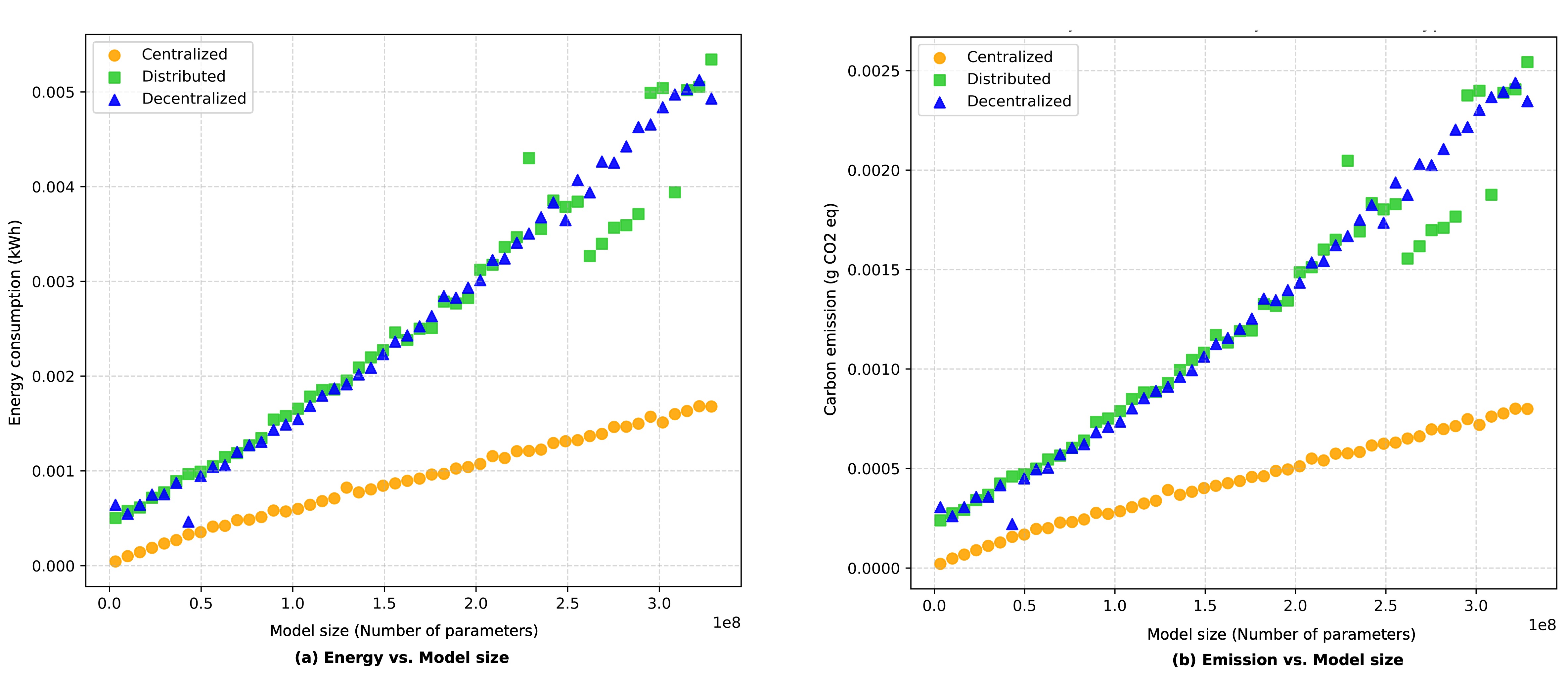}
\caption{Energy consumption and carbon emission with varying model sizes, characterized by the total number of parameters of all agents' neural networks.}
\label{f:size}
\end{figure*}

\textbf{Discussion:} As model size increases, all three models show elevated energy consumption, with the distributed and decentralized setups experiencing a more rapid escalation compared to the centralized configuration.

We conducted a comparative analysis of the energy consumption and carbon emissions of models with varying sizes, characterized by the number of parameters. The experimental setup involved a community where each agent managed 100 generators and 100 loads. To explore the impact of model size on energy consumption, the number of hidden nodes was varied to create models of different sizes. The results, illustrated in Figure \ref{f:size}, revealed the following insights:

\textbf{Energy Consumption Growth:} As shown in Figure \ref{f:size}, all three models demonstrated increased energy consumption with growing model size. This trend aligns with intuitive expectations, as larger models with more parameters necessitate greater computational resources for data processing and parameter updates. The increasing complexity of the network, driven by more hidden nodes, leads to a higher computational load and consequently higher energy use.

\textbf{Sensitivity to Model Size:} The energy consumption in distributed and decentralized setups showed a more rapid escalation compared to centralized configurations as the model size increased. This trend highlights the importance of optimizing model sizes not only for computational efficiency but also for energy efficiency, particularly in large-scale applications such as power grid management. This sensitivity suggests that in scenarios where energy efficiency is critical, careful consideration should be given to the balance between model complexity and operational feasibility.

\subsection{Results on Different System Scalability}

\textbf{Discussion:} The energy usage increase of the distributed and decentralized models is less sensitive to system scalability than the centralized model.

We compared model scalability, characterized by the number of decision-making units, i.e., \(N_{\texttt{A}}N_{\texttt{G}}\), adjusting it through different numbers of generators \(N_{\texttt{G}}\). The number of hidden nodes remained fixed in the experiments.

\begin{figure}[htbp]
\centering
\setlength{\abovecaptionskip}{0.2cm}
\includegraphics[width=1\columnwidth]{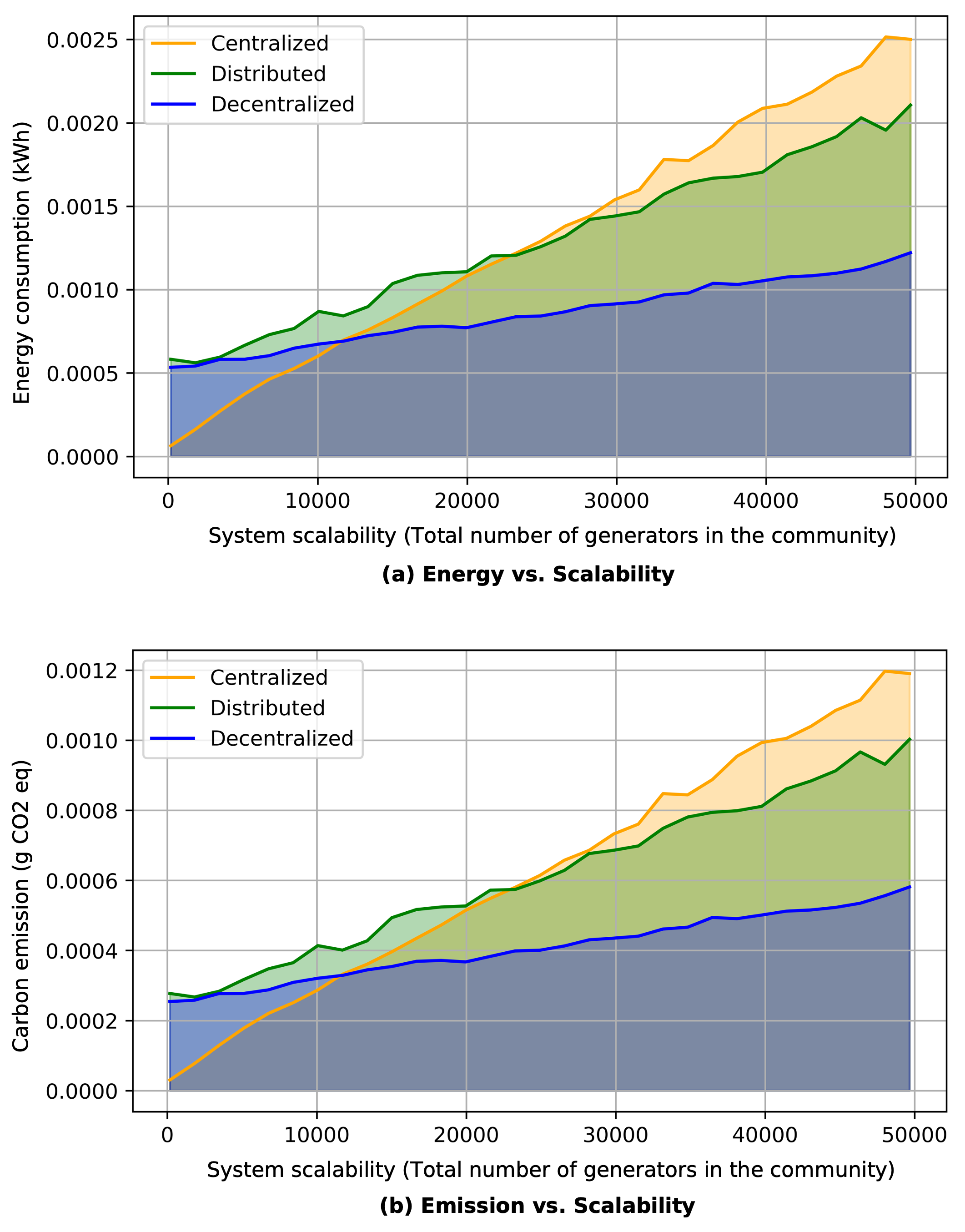}
\caption{Comparison of energy consumption and carbon emissions based on system scalability, indicated by the number of decision-making units or generators. Distributed and decentralized models exhibit less sensitivity to scalability changes in energy usage compared to centralized models.}
\label{Ng}
\end{figure}

\begin{figure}[htbp]
\centering
\setlength{\abovecaptionskip}{0.2cm}
\includegraphics[width=1\columnwidth]{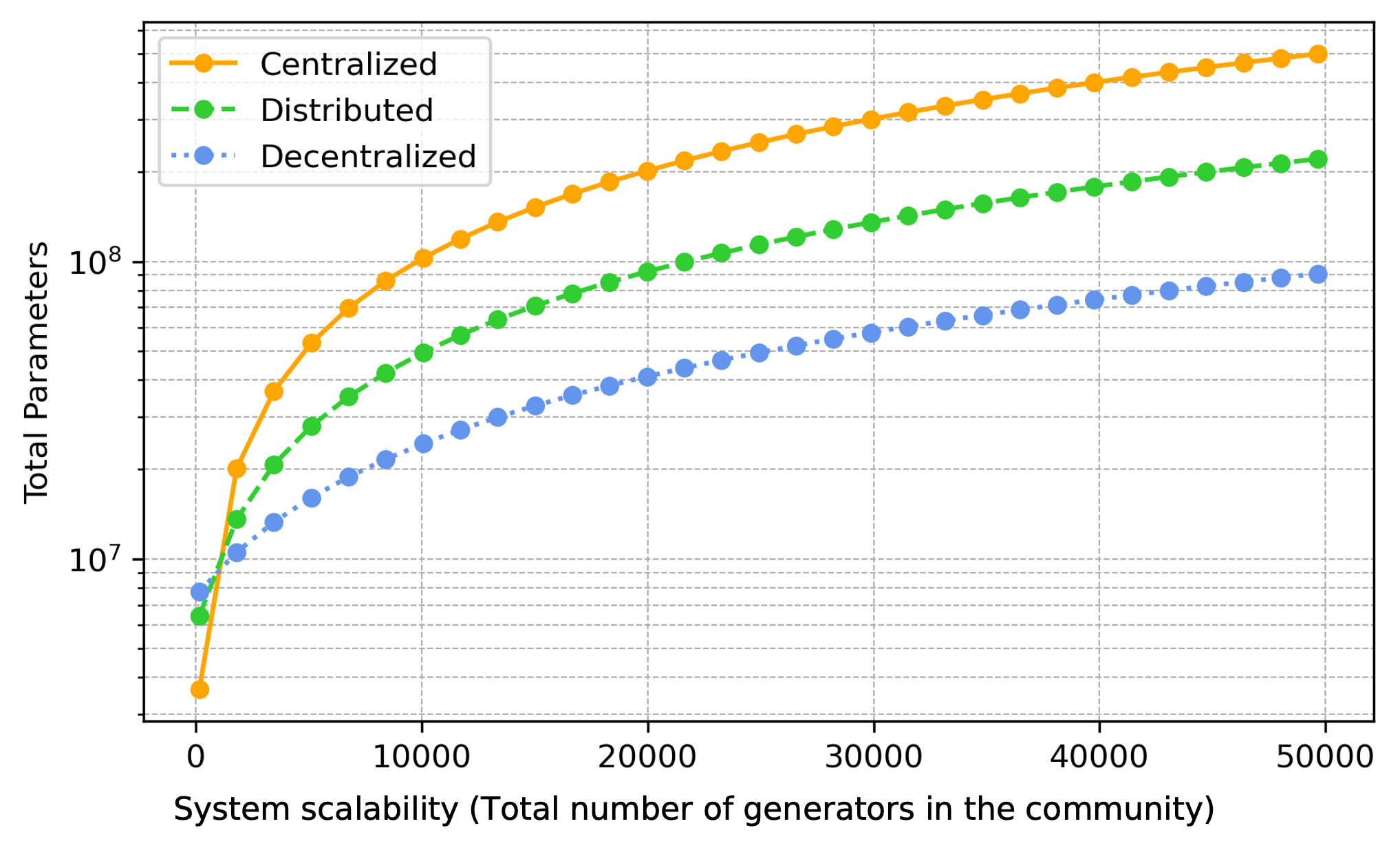}
\caption{Model size increase as the system scalability increases. The rate of increase in model size for the distributed and decentralized models is slower than for the centralized model as system scalability increases.}
\label{Ng_para}
\end{figure}

Figure \ref{Ng} illustrates how energy consumption varies as the number of decision-making units changes within the community. All models exhibit increased energy consumption as the number of generators grows, which is intuitive because more decision-making units lead to greater computational demands. However, when the number of generators is relatively small, multi-agent models (distributed and decentralized) generally consume more energy than the centralized model. As the number of generators in the community increases, the centralized model's energy consumption surpasses that of the distributed and decentralized models. This pattern may be attributed to the fact that the model size in the centralized setup increases much faster than in multi-agent models as the number of generators increases. As shown in Figure \ref{Ng_para}, the total number of parameters grows more rapidly for the centralized model than for the multi-agent models.

\newpage 
This analysis highlights the advantage of multi-agent models as the number of components in the community increases, because the total number of parameters in multi-agent models is less sensitive to increases than in the centralized model.

\section{Conclusion} 

This paper presents a pioneering comparison of the energy demands and carbon footprints associated with ML-driven power dispatch models across centralized, distributed, and decentralized configurations. Our findings reveal that distributed and decentralized setups have higher energy consumption per communication iteration and experience more rapid increases in energy usage as model size grows, compared to centralized configurations. However, these models exhibit less sensitivity to system scalability changes. This analysis highlights crucial trade-offs between operational efficiency and environmental impact, offering vital insights for future AI applications in energy systems. By exposing the significant energy and carbon costs associated with AI in power dispatch, this research advances our understanding of AI's environmental implications and guides future AI development to balance system efficiency with ecological integrity. This ensures that the evolution of power systems remains both innovative and sustainable.

\bibliographystyle{ieeetr}
\bibliography{main}

\begin{thebibliography}{10}

\bibitem{kargarian2016toward}
A.~Kargarian, J.~Mohammadi, J.~Guo, S.~Chakrabarti, M.~Barati, G.~Hug, S.~Kar, and R.~Baldick, ``Toward distributed/decentralized dc optimal power flow implementation in future electric power systems,'' {\em IEEE Transactions on Smart Grid}, vol.~9, no.~4, pp.~2574--2594, 2016.

\bibitem{li2023hard}
M.~Li, S.~Kolouri, and J.~Mohammadi, ``Learning to solve optimization problems with hard linear constraints,'' {\em IEEE Access}, vol.~11, pp.~59995--60004, 2023.

\bibitem{amos2022meta}
B.~Amos, S.~Cohen, G.~Luise, and I.~Redko, ``Meta optimal transport,'' {\em arXiv preprint arXiv:2206.05262}, 2022.

\bibitem{zhu2023intelligent}
S.~Zhu, T.~Yu, T.~Xu, H.~Chen, S.~Dustdar, S.~Gigan, D.~Gunduz, E.~Hossain, Y.~Jin, F.~Lin, {\em et~al.}, ``Intelligent computing: The latest advances, challenges, and future,'' {\em Intelligent Computing}, vol.~2, p.~0006, 2023.

\bibitem{avelar2023ai}
V.~Avelar, P.~Donovan, P.~Lin, W.~Torell, and M.~A. Torres~Arango, ``The ai disruption: Challenges and guidance for data center design,'' White Paper 110, Schneider Electric – Energy Management Research Center, 2023.
\newblock Accessed: 2024-06-24.

\bibitem{schmidt2021codecarbon}
V.~Schmidt, K.~Goyal, A.~Joshi, B.~Feld, L.~Conell, N.~Laskaris, D.~Blank, J.~Wilson, S.~Friedler, and S.~Luccioni, ``Codecarbon: estimate and track carbon emissions from machine learning computing,'' {\em Cited on}, vol.~20, 2021.

\bibitem{faiz2023llmcarbon}
A.~Faiz, S.~Kaneda, R.~Wang, R.~Osi, P.~Sharma, F.~Chen, and L.~Jiang, ``Llmcarbon: Modeling the end-to-end carbon footprint of large language models,'' {\em arXiv preprint arXiv:2309.14393}, 2023.

\bibitem{lacoste2019quantifying}
A.~Lacoste, A.~Luccioni, V.~Schmidt, and T.~Dandres, ``Quantifying the carbon emissions of machine learning,'' {\em arXiv preprint arXiv:1910.09700}, 2019.

\bibitem{luccioni2023counting}
A.~S. Luccioni and A.~Hernandez-Garcia, ``Counting carbon: A survey of factors influencing the emissions of machine learning,'' {\em arXiv preprint arXiv:2302.08476}, 2023.

\bibitem{luccioni2023power}
A.~S. Luccioni, Y.~Jernite, and E.~Strubell, ``Power hungry processing: Watts driving the cost of ai deployment?,'' {\em arXiv preprint arXiv:2311.16863}, 2023.

\bibitem{molzahn2017survey}
D.~K. Molzahn, F.~D{\"o}rfler, H.~Sandberg, S.~H. Low, S.~Chakrabarti, R.~Baldick, and J.~Lavaei, ``A survey of distributed optimization and control algorithms for electric power systems,'' {\em IEEE Transactions on Smart Grid}, vol.~8, no.~6, pp.~2941--2962, 2017.

\bibitem{li2023machine}
M.~Li and J.~Mohammadi, ``Machine learning infused distributed optimization for coordinating virtual power plant assets,'' {\em arXiv preprint arXiv:2310.17882}, 2023.

\bibitem{li2023learning}
M.~Li, S.~Kolouri, and J.~Mohammadi, ``Learning to optimize distributed optimization: Admm-based dc-opf case study,'' in {\em 2023 IEEE Power \& Energy Society General Meeting (PESGM)}, pp.~1--5, IEEE, 2023.

\bibitem{biagioni2020learning}
D.~Biagioni, P.~Graf, X.~Zhang, A.~S. Zamzam, K.~Baker, and J.~King, ``Learning-accelerated admm for distributed dc optimal power flow,'' {\em IEEE Control Systems Letters}, vol.~6, pp.~1--6, 2020.

\bibitem{mak2023learning}
T.~W. Mak, M.~Chatzos, M.~Tanneau, and P.~Van~Hentenryck, ``Learning regionally decentralized ac optimal power flows with admm,'' {\em IEEE Transactions on Smart Grid}, 2023.

\bibitem{zimmerman2010matpower}
R.~D. Zimmerman, C.~E. Murillo-S{\'a}nchez, and R.~J. Thomas, ``Matpower: Steady-state operations, planning, and analysis tools for power systems research and education,'' {\em IEEE Transactions on power systems}, vol.~26, no.~1, pp.~12--19, 2010.

\end{thebibliography}

\end{document}